\documentclass[sigconf]{acmart}
\usepackage{multirow,tabularx} 
\usepackage{booktabs}
\usepackage{color, colortbl} 
\usepackage{array, booktabs, makecell}
\usepackage{comment} 
\usepackage{color,soul}
\usepackage{graphicx}
\usepackage{caption}
\usepackage{subcaption}
\usepackage{nth}
\usepackage{url}
\usepackage{textcomp}
\usepackage{pgfplots}
\usepackage{tikz}
\usetikzlibrary{patterns}
\usepackage[figuresright]{rotating}
\usepackage{textcomp}  
\usepackage{mathtools}
\usepackage{amsmath}
\usepackage{tabularx}
\usepackage{pgfplots}
\usepackage{pgfplotstable}
\usepackage{sfmath}
\usepackage{array, booktabs, makecell}
\usepackage{color}
\usepackage{csquotes}
\usepackage{url}
\usepackage{comment}
\usepackage{tikz}
\usepackage{listings}
\usepackage{booktabs} 
\usepackage{multirow}
\usetikzlibrary{fit} 
\usetikzlibrary{positioning}
\usetikzlibrary{arrows}
\usetikzlibrary{shapes.multipart}
\usepackage{float}
\usepackage{enumitem}
\usepackage{mathtools}
\usepackage{pgfplots}
\usepgfplotslibrary{fillbetween}
\pgfplotsset{compat=1.11,width=10cm}
\usepgfplotslibrary{statistics}

\usepackage{pgfplots, pgfplotstable}

\usepackage{bchart}
\usepackage{fancybox}
\usepackage{pgf-pie}  
\usepackage{tabularx}
\usepackage{pgfplots}
\usepackage{pifont}
\usepackage{wrapfig}

\usetikzlibrary{decorations.pathmorphing}
\tikzset{snake it/.style={decorate, decoration=snake}}
\usepackage{xcolor}
\definecolor{MyOrange}{RGB}{237,125,49}
\definecolor{MyBlue}{RGB}{91,155,213}
\title{Skill, Will, or Both? Understanding Digital Inaccessibility from Accessibility Professionals’ Viewpoint} 

\author{P D Parthasarathy}
\email{p20210042@goa.bits-pilani.ac.in}
\orcid{0000-0002-8723-2407}
\affiliation{%
\institution{BITS Pilani, KK Birla Goa Campus, Goa, India}
\city{}\state{}
\country{}
\postcode{}
}

\author{Rachel F. Adler}
\email{radler@illinois.edu}
\orcid{0000-0002-0462-5716}
\affiliation{%
\institution{\hbox{University of Illinois Urbana-Champaign,} Illinois, USA}
\city{}\state{}
\country{}
\postcode{}
}

\author{Devorah Kletenik}
\email{kletenik@sci.brooklyn.cuny.edu}
\orcid{0000-0003-4362-3884}
\affiliation{%
\institution{Brooklyn College, City University of New York, New York, USA}
\city{}\state{}
\country{}
\postcode{}
}

\author{Swaroop Joshi}
\email{swaroopj@goa.bits-pilani.ac.in}
\orcid{0000-0003-4536-2446}
\affiliation{%
\institution{BITS Pilani, KK Birla Goa Campus, Goa, India}
\city{}\state{}
\country{}
\postcode{403726}
}

\author{Anshu M Mittal}
\email{f20231125@goa.bits-pilani.ac.in}
\orcid{0009-0004-4481-8645}
\affiliation{%
\institution{BITS Pilani, KK Birla Goa Campus, Goa, India}
\city{}\state{}
\country{}
\postcode{403726}
}

\definecolor{light-gray}{rgb}{0.95,0.95,0.95}

\usepackage{tikz}

\copyrightyear{2025}
\acmYear{2025}
\setcopyright{rightsretained}
\acmConference[CHI EA '25]{Extended Abstracts of the CHI Conference on Human Factors in Computing Systems}{April 26-May 1, 2025}{Yokohama, Japan}
\acmBooktitle{Extended Abstracts of the CHI Conference on Human Factors in Computing Systems (CHI EA '25), April 26-May 1, 2025, Yokohama, Japan}\acmDOI{10.1145/3706599.3720277}
\acmISBN{979-8-4007-1395-8/2025/04}

\begin{document}

\begin{abstract}
Digital inaccessibility continues to be a significant barrier to true inclusion and equality. WebAIM's 2024 report reveals that only 4.1\% of the world's top one million website homepages are fully accessible. Furthermore, the percentage of web pages with detectable Web Content Accessibility Guidelines (WCAG) failures has only decreased by 1.9\% over the past five years, from 97.8\%. To gain deeper insights into the persistent challenges of digital accessibility, we conducted a comprehensive survey with 160 accessibility professionals. Unlike previous studies, which often focused on technology professionals, our research examines inaccessibility through the lens of dedicated accessibility professionals, offering a more detailed analysis of the barriers they face. Our investigation explores (a) organizations' willingness to prioritize accessibility, (b) the challenges in ensuring accessibility, and (c) the current accessibility training practices in technology workspaces. This study aims to provide an updated perspective on the state of digital accessibility from the point of view of accessibility professionals.
\end{abstract}


\maketitle


\section{Introduction}
\label{sec:intro}
Digital accessibility\footnote{We use accessibility or a11y to refer to digital accessibility throughout this paper.} refers to the process of designing and developing digital artifacts, such as websites, applications, and digital content, in a way that makes them usable by \emph{everyone}, including those with disabilities. The goal is to provide equitable access, ensuring that people with disabilities can engage with and benefit from digital resources on par with those without disabilities. Far from being an optional consideration, digital accessibility is an essential requirement of all software, critical to fostering inclusivity in the digital landscape.

Accessibility is mandated by various governments, including the United States~\cite{the_us_government_it_2023}, European Union~\cite{europian_commission_web_2023}, Canada~\cite{secretariat_standard_2011}, and India~\cite{ministry_of_social_justice_and_empowerment_government_of_india_rights_2016}. However, according to the 2024 WebAIM report~\cite{webaimWebAIMMillion2024}, only 4.1\% of the top 1 million websites provide full accessibility. Meanwhile, nearly 97\% of around 500 popular Android apps in 23 business categories show accessibility violations~\cite{yan_current_2019}.
The W3C created the Web Content Accessibility Guidelines (WCAG) in 1999 and periodically updates them, with version 2.2 being the latest, to promote the creation of accessible digital products and services; ``yet, even after 25 years, most pages still don't meet these guidelines well''~\cite{branham_state_2024}. As per the 2021 UserWay report~\cite{userway_economic_2021}, inaccessible applications and websites cause a global revenue loss of around \$16 billion in the e-commerce sector alone. This highlights that inaccessibility is not merely a social justice issue but also a significant driver of financial losses and legal risks for businesses. Both the W3C and existing literature recommend incorporating accessibility considerations early in the software development lifecycle ~\cite{w3c_start_2014}. The literature also highlights the essential role of human involvement—particularly software professionals—in addressing accessibility challenges. This emphasis is well-founded, as ensuring accessibility is a shared responsibility. Software designers are tasked with creating accessible user experiences, while software engineers ensure accessibility by writing compliant code, interpreting results from evaluation tools, and addressing any identified accessibility violations. Software managers play a key role in hiring professionals with accessibility expertise, allocating resources for tasks like testing and remediation, and prioritizing accessibility within the development process. Unfortunately, the existing university-level computing education does not adequately prepare students to address accessibility topics~\cite{patel_why_2020, parthasarathy_exploring_2024, shinohara_who_2018, soares_guedes_how_2020, parthasarathy_teaching_2024}. Consequently, to bridge this accessibility skill gap, organizations often employ dedicated accessibility subject matter experts (SMEs) or consultants who take responsibility for training teams in accessibility skills, overseeing the design and development of accessible software, generating accessibility conformance reports, and fostering a culture and mindset of inclusive design. Despite a growing number of professionals obtaining accessibility certifications and an increasing demand for accessibility SMEs ~\cite{levelaccessStateDigitalAccessibility2024}, the overall state of digital accessibility has shown limited improvement. While the existing literature acknowledges a skill gap as a barrier ~\cite{patel_why_2020, parthasarathy_exploring_2024, coverdale_digital_2024, coverdale_teaching_2022, lewthwaite_researching_2023}, it raises the question of whether other factors, such as organizational willingness, contribute to the persistent challenges. This study seeks to explore these barriers by gathering insights from accessibility professionals to better understand the factors beyond skill that may hinder progress toward more inclusive digital experiences. Our research questions (RQs) are:

\begin{description}
  \item [RQ1] To what extent are organizations willing to prioritize accessibility?
  \item [RQ2] What challenges do organizations face in ensuring accessibility despite having dedicated accessibility roles?
  \item [RQ3] How is accessibility training implemented, and how effective are these training programs?
\end{description}

This paper is organized as follows: Sec.~\ref{sec:RelatedWork} describes the related literature, and Sec.~\ref{sec:StudyDesign} explains the study design in detail. The findings are presented in Sec.~\ref{sec:StudyResults} and discussed in Sec.~\ref{sec:takeaway} including the limitations and future work, before concluding in Sec.~\ref{sec:conc}.
\section{Related Work}
\label{sec:RelatedWork}

Several studies suggest that formal education frequently falls short of equipping professionals with sufficient accessibility competencies~\cite{the_partnership_on_employment__accessible_technology_peat_accessible_2018, parthasarathy_exploring_2024, coverdale_digital_2024, coverdale_teaching_2022, lewthwaite_researching_2023}. While accessibility education at the university level is expanding, training for IT professionals remains limited. Most job postings for software designers and developers in 2021 did not mention accessibility~\cite{martin_landscape_2022}, and a workplace survey that year indicated accessibility was not a company priority~\cite{miranda_studying_2022}.

Similarly, in an earlier study, Lazar et al.~\cite{lazar_improving_2004} found in 2004 that while many webmasters supported accessibility in principle, practical barriers such as limited time, training, and managerial support led to inaccessible websites. In 2018-2019, interviews with 30 accessibility practitioners from 13 companies found a need to educate their co-workers about accessibility since it was lacking in their education~\cite{azenkot_how_2021}. Further, interviewees felt that current accessibility guidelines were hard to understand and the most effective format would be training sessions with new employees. In 2020, \citeauthor{patel_why_2020}~\cite{patel_why_2020} surveyed 77 technology professionals from the United States to investigate their perspectives and challenges related to accessibility. The study revealed gaps in formal accessibility expertise, limited availability of tools and resources, and a failure to account for retroactive accessibility adjustments within project timelines. A similar study conducted in India in 2023 with 269 participants~\cite{parthasarathy_exploring_2024} also highlighted a lack of formal accessibility expertise and insufficient tools and resources among professionals. WebAIM's 2021 survey of 758 web accessibility practitioners globally~\cite{webaim_survey_2021} found that nearly half of the respondents acknowledged their organization's web products were not highly accessible. Around 40\% reported no significant improvement in accessibility compared to previous years, while 13\% observed a decline in accessibility standards. \citeauthor{bi_accessibility_2022} \cite{bi_accessibility_2022} surveyed professionals across 26 countries, revealing a widespread lack of accessibility expertise and the need for executive support and technical training. 

In a 2019 publication, ~\citeauthor{crabb_developing_2019} ~\cite{crabb_developing_2019} report on findings of a workshop which showed that there are areas ``where developers struggle to empathize with accessibility issues and subsequently design interactions for this demographic,'' as well as a ``lack of understanding in how a person with disability uses technology'' and how that impacts ``how technology interactions are designed.'' A 2023 report by Teach Access with 107 tech leaders presents a significant accessibility skills gap: 56\% of organizations struggle to find candidates with accessibility expertise, while only 2\% find it easy. Additionally, 44\% reported inadequate accessibility skills among current staff, with 75\% noting increased demand for these skills in the past five years and 86\% expecting this demand to grow ~\cite{teach_access_accessible_nodate}. There is also an increased emphasis on \emph{shift-left} approach to accessibility ~\cite{armas_proposal_2020, parthasarathy_exploring_2024}. This approach to accessibility emphasizes incorporating accessibility considerations early in the design and development phases rather than postponing them to later stages, where addressing issues becomes more time-consuming and resource-intensive. This approach promotes the idea that accessibility is a shared responsibility across all roles—designers, developers, and management—rather than being solely the responsibility of testers or accessibility SMEs. While the term \textit{shift-left} has gained popularity in recent years, the underlying concept has been advocated since at least 2005 ~\cite{law_programmer-focused_2005}. However, a 2022 study of LinkedIn job posts found that the majority of accessibility-related posts focused on accessibility testing rather than accessibility integration into earlier stages of development, thus continuing to view accessibility as an afterthought~\cite{martin_landscape_2022}.


Limited research has explored the influence of management support or examined accessibility challenges from the perspective of accessibility professionals. Our study aims to address this gap by investigating whether, beyond skill gaps, the willingness of organizational leadership contributes to digital inaccessibility and the barriers to ensuring accessibility, as perceived by accessibility practitioners.

\section{Method}
\label{sec:StudyDesign}
This section outlines our study design and the approaches used for participant recruitment.

\subsection{Survey Design}
An online survey was designed and conducted from mid-July to mid-December 2024 to explore our research questions. The survey began with a brief explanation of digital accessibility, followed by an informed consent form approved by the institute's Human Ethical Committee. Only participants who provided consent could proceed. The second question confirmed whether the respondent currently or previously held an accessibility-related role, with only those meeting this criterion continuing further. Several survey questions were adapted from validated instruments used in previous research \cite{shinohara_who_2018, parthasarathy_exploring_2024}, while others were developed based on our expertise in the field. The core survey questions are included in the Appendix for reference. The survey was conducted using Zoho Forms\footnote{\url{https://www.zoho.com/forms/}}, chosen for its ability to capture partial responses and support advanced form controls, such as star rating questions.

\subsection{Recruiting Participants}
To explore the research questions (RQs) from the perspective of accessibility professionals, we employed purposive sampling to ensure the participation of individuals with relevant expertise. The online survey was distributed through multiple channels to maximize reach and diversity among respondents: 

\begin{itemize}[leftmargin=*]
    \item \textbf{Email Invitations:} Participants were recruited by sending email invitations to a pool of 73 professionals within the software industry, selected based on their relevance to the study’s focus area.
    \item \textbf{Accessibility Community Groups:} The survey link was shared in various professional accessibility forums. Shared the survey in a global accessibility Slack community (1,470 members), a WhatsApp group for Global Accessibility Consultants (251 members), and the X Accessibility Community (723 members). 
    \item \textbf{Accessibility Events and Conferences:} The survey was promoted at conferences and accessibility-focused events held during the data collection period to engage a broader audience of accessibility professionals.
\end{itemize}

\subsection{Data Analysis}
The survey responses collected via Zoho Forms were exported in CSV format and securely stored in restricted-access cloud storage for analysis. The data was subsequently analyzed using statistical techniques in Python. 
\section{Findings}
\label{sec:StudyResults}
A total of 165 responses were collected. However, five respondents indicated in the second survey question that they were not in an accessibility-related role and, therefore, did not proceed further. These responses were excluded from the analysis, leaving 160 complete responses for consideration. 

\subsection{Participants' Profile}
A diverse set of accessibility professionals participated in the survey. Some highlights are that the participant pool was highly experienced, with nearly half (48.1\%) having over 8 years of accessibility. Most held specialized roles, including Accessibility SMEs/Consultants (56.8\%), Testers (15.6\%), and Managers (8.8\%). Geographically, participants were primarily from North America (38.8\%), Europe (29.3\%), and Asia (22.5\%). Gender representation included 48.1\% males and 33.8\% females. Notably, 27.5\% identified as persons with disabilities, citing conditions such as low vision (26.5\%), blindness (9.2\%), autism (11.2\%), mental health-related conditions (17.3\%), and hearing loss (9.2\%).

Table \ref{tab:participantSummary} (in Appendix) details the participants' profiles. Regarding educational background, 31.9\% of respondents had qualifications in computer science or related fields,  8.7\% came from UX and design disciplines, and the remaining 59.4\% had diverse academic backgrounds, including biology, arts, history, literature, and journalism. When assessing their proficiency in digital accessibility, including the WCAG, 43.12\% of respondents rated themselves as \emph{Experts} (5 out of 5 stars), while 43.75\% identified as \emph{Proficient} (4 out of 5 stars). The remaining 13.12\% rated themselves at an \emph{Intermediate} level (3 out of 5 stars). The respondents' average ratings amounted to 4.3, with a standard deviation of 0.68.

The participants of the survey represented a diverse group of organization types. Over half (56.2\%) of the respondents work in large-scale companies (over 1000 employees), while around a third (36.9\%) belong to medium-scale organizations (100-1000 employees). Nearly half (48.1\%) are employed in Business-to-Business (B2B) organizations (firms that develop and transact with other business entities than directly with consumers), while around 40.6\% work in Business-to-Consumer (B2C) companies (firms which directly caters to consumers via their products). Approximately a third (33.7\%) work in product-based companies, while around a quarter (27.5\%) are in service-based firms. Notably, 32.5\% work in hybrid setups combining both product \& service models. Only 6.3\% belong to other organizational types such as non-profit, library, and education services. Table~\ref{tab:participantOrgSummary} (in Appendix) describes the respondent's organization-related findings in detail.

\subsection{RQ1: Organizations Willingness to Prioritize Accessibility}

An \emph{accessibility champion} is a trained accessibility expert within an organization who advocates for accessibility, serves as a point of contact, and supports their team on accessibility-related topics \cite{azenkot_how_2021}. 41.9\% of the participants (N=160) reported that their organization lacked at least one accessibility champion per team with expertise in accessibility topics. Meanwhile, 30.6\% were unsure if such champions existed within their organization, and only 27.5\% confirmed having an accessibility champion skilled in accessibility topics in each team to provide support.

\begin{figure}[h]
  \centering
  \begin{minipage}[b]{0.49\textwidth}
      \centering
    \includegraphics[width=0.7\textwidth]{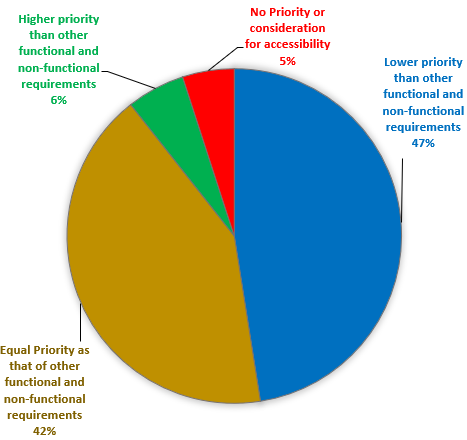}
    \caption{Indicates a11y prioritization by organizational leadership (N=160)}
    \Description{Indicates accessibility prioritization with 47\% feeling a11y is a lesser priority than other requirements.}
    \label{fig:prioritization}
  \end{minipage}
  \hfill
  \begin{minipage}[b]{0.49\textwidth}
    \centering
    \includegraphics[width=0.66\textwidth]{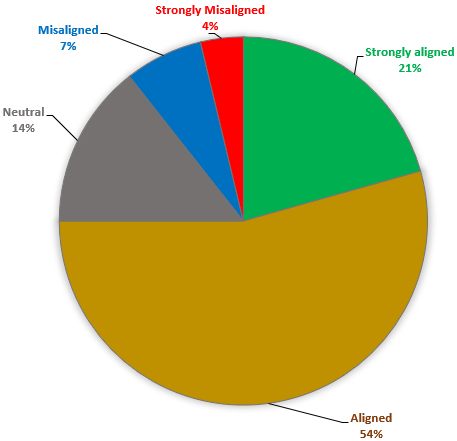}
    \caption{Indicates alignment of organizational values with a11y goals (N=160)}
    \Description{Indicates alignment of organizational values with accessibility goals where only 21\% are strongly aligned }
    \label{fig:orgValues}
  \end{minipage}
\end{figure}

Fig \ref{fig:prioritization} shows the prioritization of digital accessibility by organizational leadership (engineering managers and upper management) compared to other functionalities; nearly half (47\%, N=160) felt that accessibility was given lower priority than other functional and non-functional requirements, and 5\% indicated that accessibility did not receive priority or consideration. In contrast, 42\% reported that their organizations prioritized accessibility equally with other requirements. Finally, 6\% noted that accessibility was prioritized above other requirements. This last group comprised nine respondents, of which 8 belonged to a B2B, service-based organization and one belonged to a B2C, hybrid-based organization. 

Also, as shown in Fig \ref{fig:orgValues}, a small portion of participants (4\%) felt that their organization's values and missions were strongly misaligned with accessibility goals, while 7\% reported it being misaligned, and 14\% remained neutral. Over half (54\%) indicated that their organization's values and missions were aligned with accessibility goals, whereas only 21\% noted a strong alignment. Over 80\% of the participants who reported a strong alignment between their organization's values and accessibility goals were from business-to-business (B2B) companies, where stringent legal requirements for software accessibility often influence organizational priorities when selling to other businesses.

A common initial strategy for companies committed to accessibility involves introducing initiatives and incentives, such as Global Accessibility Awareness Day celebrations and accessibility bug bounty programs (which is a dedicated time to identify and fix accessibility bugs)~\cite{parthasarathy_teaching_2024-1, azenkot_how_2021}. However, our findings revealed that 68.1\% of participants (N=160) reported that their organizations had no such initiatives or incentives to motivate engineers and designers to prioritize accessibility. In contrast, 21.3\% indicated their organizations had implemented such efforts, while 10.6\% noted that initiatives were currently in progress. Participants who reported having initiatives in place mentioned strategies such as recognizing and rewarding teams for developing accessible products, offering prizes, hosting hackathons and bug bounties, and organizing celebratory events. A few also highlighted the implementation of a champions network and incorporating accessibility performance metrics into employee evaluations.

Organizations that excel in accessibility typically comply with WCAG, set metrics to measure progress, foster collaboration, allocate budgets for training and enforce mandatory accessibility reviews before product releases. We explored these aspects, and the findings are presented in Table \ref{tab:establishedprocesses}.

\begin{table}[h]
  \centering
  \caption{Accessibility Commitment with established metrics, training budget \& quality reviews (N=160)} \label{tab:establishedprocesses}
  \begin{tabular}{p{4cm}rrr}
    \toprule
     Metric    & Yes  & No & Not Sure \\ \midrule
     Established metrics to measure the success of accessibility efforts & 25.6\%   & 74.4\% & 0.00\% \\
     Dedicated budget or resources for digital accessibility training and initiatives  & 15.0\% &  59.4\%  & 25.6\%  \\
     Mandatory quality reviews before product release by skilled accessibility professionals  & 48.1\%  & 51.9\% & 0.00\% \\  
     \bottomrule 
  \end{tabular}
\end{table}

\textbf{\textit{RQ1 Answer:}} Many organizations still lack commitment to accessibility. Nearly half (47\%) are perceived as prioritizing it lower than other requirements, and just 21\% strongly align their mission with accessibility goals. Most lack initiatives or incentives (68.1\%), and few have success metrics (25.6\%), training budgets (15\%), or mandatory reviews (48.1\%), highlighting insufficient leadership and structural support for accessibility.


\subsection{RQ2: Challenges in Ensuring Accessibility Despite Dedicated Roles}


Participants identified the top barriers to ensuring digital accessibility (multi-select question) as follows: lack of awareness (68.75\%), inadequate prioritization leading to insufficient time (66.87\%), perceptions of low return on investment (55.62\%), unwillingness to address accessibility despite available resources (53.12\%), challenges enforcing the “shift-left” principle (48.12\%), shortage of skilled professionals (47.5\%), and budgetary constraints (46.25\%).

When asked about the stage at which accessibility is considered in the product development lifecycle, 58\% of them defer accessibility to later stages of engineering and testing phases and only 10.5\% reported addressing it starting from the User Experience (UX) research phase, while 31.5\% indicated incorporating it starting from the UX design phase. Key challenges in adopting shift-left approach, as reported by participants, include limited engagement (68.75\%), lack of management support (53.12\%), bureaucratic hurdles around accountability (47.5\%), lack of awareness (32.5\%), and organizational structures that hinder shift-left adoption (28.12\%). Additionally, some respondents highlighted other barriers, such as the absence of standardized collaboration practices, pressure for faster output, and limited time for accessibility training and implementation.

\begin{figure}[h]
  \centering
  \begin{minipage}[b]{0.49\textwidth}
    \centering
    \includegraphics[width=0.65\textwidth]{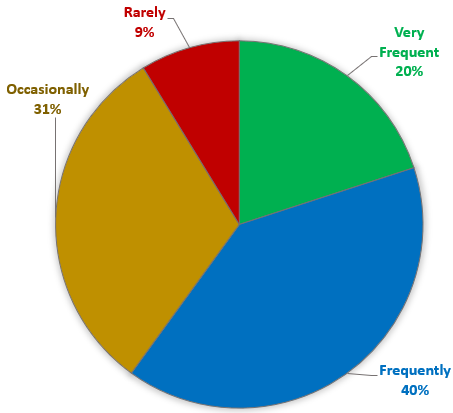}
    \caption{Indicates collaboration frequency of accessibility SMEs with other designers and engineers (N=160)}
    \Description{Indicates collaboration frequency of accessibility SMEs with other designers and engineers }
    \label{fig:collab}
  \end{minipage}
  \hfill
  \begin{minipage}[b]{0.49\textwidth}
    \centering
    \includegraphics[width=0.65\textwidth]{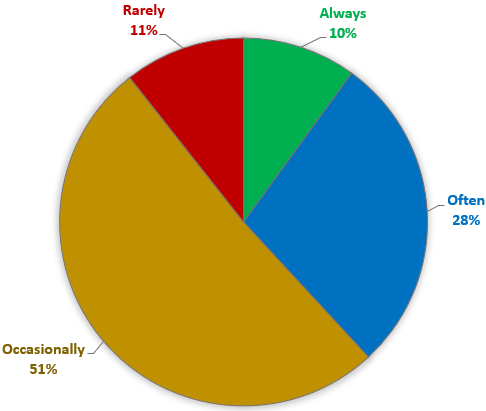}
    \caption{Indicates the rate of incorporation of accessibility feedback received from accessibility SMEs (N=160)}
    \Description{Indicates rate of incorporation of accessibility feedback received from accessibility SMEs}
    \label{fig:feedbackIncorporated}
  \end{minipage}
\end{figure}

Fig \ref{fig:collab} shows the frequency of collaboration between accessibility professionals and designers/engineers; 20\% reported very frequent collaboration, 40\% indicated frequent collaboration, 31.25\% mentioned occasional collaboration, and 8.75\% reported rare collaboration. Notably, 78\% of those experiencing very frequent collaboration belong to business-to-business (B2B) organizations, and 65\% are part of large-sized organizations.

Fig 4\ref{fig:feedbackIncorporated} depicts the frequency with which designers and engineers incorporate feedback provided by accessibility experts. Only 10\% reported \textit{Always} incorporating feedback, while 28.12\% said they do so \textit{Often}. In contrast, over half (51.25\%) admitted to incorporating feedback only \textit{Occasionally}, and 10.62\% reported doing so \textit{Rarely}. 

\textbf{\textit{RQ2 Answer:}} Key barriers to digital accessibility include lack of awareness, low prioritization, reluctance to act despite available resources, and budget constraints. Accessibility is rarely integrated early due to obstacles such as limited engagement, lack of management support, and bureaucratic hurdles. Collaboration between accessibility experts and designers/engineers is infrequent, with over half incorporating accessibility feedback only occasionally.


\subsection{RQ3: Accessibility Training and its Effectiveness }


Out of 160 respondents, 148 (92.5\%) reported conducting accessibility training for designers, engineers, and management. Of these, 23.65\% focused solely on raising awareness and sensitizing participants about accessibility, 11.49\% concentrated exclusively on enabling the creation of accessible products by training on WCAG and inclusive design practices, and 64.86\% addressed both aspects. A significant portion (41.89\%) reported that their training sessions lasted less than 3 hours annually. About 27.70\% conducted sessions lasting 3 to 5 hours, 14.18\% provided 5 to 10 hours of training, and 16.21\% held trainings exceeding 10 hours per year. Of those offering training exceeding 10 hours, 83.33\% were large organizations, and 75\% operated under a B2B model.

\begin{figure}[h]
    \centering
    \includegraphics[width=0.9\linewidth]{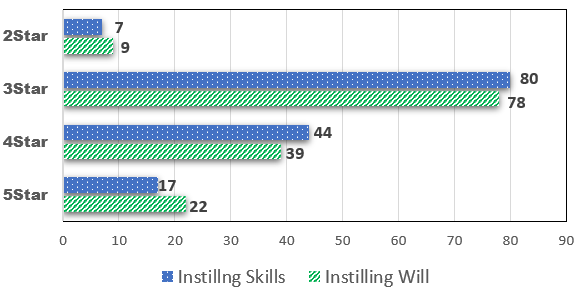}
    \caption{Comparison of effectiveness in instilling skill and will in accessibility trainings with N=148}
      \Description{Comparison of effectiveness in instilling skill and will in accessibility trainings. 3-star ratings dominate, with the highest numbers for both instilling skills (80) and instilling will (78), indicating this is the most common perception of training effectiveness.
For 4-star ratings, instilling skills (44) slightly exceeds instilling will (39).
5-star ratings show a reverse trend, with instilling will (22) surpassing instilling skills (17), indicating a stronger emphasis on motivation over skill-building at the highest level.
The lowest ratings (2-star) show minimal differences, with instilling will (9) slightly outperforming instilling skills (7).}
  \label{fig:a11ySkillWill}
\end{figure}

Regarding training modalities (multi-select), online instructor-led sessions were the most popular, used by 79.72\%, followed by offline instructor-led sessions at 51.35\%. Online pre-recorded videos were utilized by 44.59\%, while 6.7\% hosted offline watch parties for pre-recorded content. Notably, one respondent incorporated animated videos, games, and simulations for accessibility training.
Fig \ref{fig:a11ySkillWill} provides insights into the perceived effectiveness of accessibility training in two critical areas: instilling skills (practical knowledge and techniques) and instilling will (motivation and commitment). 3-star/neutral ratings dominate for both dimensions, suggesting that while accessibility training is moderately effective overall, there is substantial room for improvement in making it highly impactful.

Over half of the respondents (53.1\%, N=160) reported that their organizations neither provide time or funding for accessibility learning nor encourage employees to pursue certifications like Section 508 Trusted Tester or IAAP\footnote{\url{https://www.accessibilityassociation.org/s/}}. A quarter of the respondents indicated organizational support for such initiatives, while 21.9\% were unsure.
Most respondents provide accessibility training, typically under 3 hours annually, with online instructor-led sessions being the most common. While training effectiveness is moderate, with motivation slightly favored over skills, more than half of organizations lack support for ongoing learning or certifications, signaling a need for stronger commitment to accessibility.

\textbf{\textit{RQ3 Answer:}} Most respondents conduct short accessibility training sessions (under 3 hours annually), focusing on awareness and skill-building, with online instructor-led formats being most common. Innovative methods like games or simulations are rare. Training effectiveness is moderate, slightly favoring motivation over skills. Over half of organizations lack support for learning or certifications, underscoring the need for stronger commitment.

\section{Discussion}
\label{sec:takeaway}

The findings from this study reflect a persistent gap in organizational commitment and implementation of digital accessibility, mirroring trends observed over the past two decades as reported by several researchers \cite{lazar_improving_2004, bi_accessibility_2022, miranda_studying_2022}. 

The findings from RQ1–RQ3 illustrate a reinforcing cycle of challenges. Organizations' low prioritization of accessibility (RQ1) creates structural gaps, such as a lack of established success metrics and insufficient budget allocation, which in turn hinder the implementation of robust training programs (RQ3). Key barriers such as insufficient prioritization (47\% reporting accessibility as a lower priority) and lack of awareness (68.75\%) reflect systemic issues in how organizations approach accessibility. For instance, the perception of low return on investment (55.62\%) likely stems from organizations viewing accessibility compliance as a cost rather than a competitive advantage. This mindset highlights the need for organizational leaders to frame accessibility as a driver of innovation and market expansion, particularly in regions with growing regulatory requirements.

This absence of systemic support exacerbates the barriers identified in RQ2, such as inadequate prioritization, limited time, and the shortage of skilled professionals. Additionally, the findings show that organizations failing to integrate accessibility early in the design process (``shift-left'') often lack the leadership commitment and collaborative frameworks necessary to sustain accessibility efforts. By linking these findings, we observe that addressing challenges in isolation may not be effective; rather, a coordinated approach is needed where organizational willingness, comprehensive training, and structural support are aligned to promote accessibility as a core business strategy.

We note that not all organizations even employ accessibility professionals in the first place.\footnote{In a recent study of over 3000 software professionals, about 77\% indicated that their workplace employed a dedicated accessibility professional~\cite{applause_state_nodate}; thus over 20\% of workplaces likely do not employ such professionals.} Organizations that do have  dedicated accessibility professionals may seem well-positioned to advance accessibility efforts, but the findings reveal that accessibility still often remains a low priority. Despite the expertise these professionals bring, systemic barriers like weak leadership support, insufficient funding, and unclear success metrics limit their impact. This paradox highlights that while employing accessibility professionals signals progress, it does not guarantee organizational readiness to treat accessibility as a core value. To bridge this gap, organizations must not only utilize these professionals’ expertise but also build a culture where accessibility is a strategic and ethical priority.

\subsection{Recommendations for Action}

\begin{itemize}[leftmargin=*]
    \item \emph{Integrate Accessibility into Organizational Culture:} Organizations need to prioritize accessibility as an integral part of user experience and product development, supported by strong leadership commitment and adequate resource allocation. Emphasizing the shift-left approach and implementing structured accessibility quality reviews with clear success metrics are essential for driving meaningful progress.
    \item \emph{Enhance Training Programs:} As recommended by \citeauthor{parthasarathy_exploring_2024}, organizations need to develop comprehensive and continuous training that includes foundational knowledge, hands-on practice, mentorship by accessibility SMEs, along with ongoing events and newsletters, to build stronger skills and ensure a lasting impact ~\cite{parthasarathy_exploring_2024}.
    \item \emph{Support Professional Development:} Offer certifications and formal career pathways to strengthen expertise.
    \item \emph{Policy Incentives:} Policymakers should introduce robust incentives like recognition programs and funding opportunities to encourage organizations to prioritize accessibility beyond compliance.
\end{itemize}

\subsection{Limitations and Future Work}

This study primarily relies on self-reported data from accessibility professionals, which may introduce biases. Additionally, the sample skews toward larger organizations and does not fully capture the perspectives of smaller firms or non-profit entities. Future research could explore these gaps by examining accessibility challenges across different organizational sizes, industries, and geographic regions. We also plan to conduct qualitative studies with participants who have consented to future research to gain deeper insights. Longitudinal studies could also track the impact of emerging training practices and organizational initiatives on accessibility outcomes.

\section{Conclusion}
\label{sec:conc}
We surveyed 160 accessibility professionals to explore persistent challenges in digital accessibility, focusing on organizational prioritization, barriers to implementation, and training practices. The results revealed significant gaps in commitment, including inadequate support, limited training, and systemic barriers. This expert-driven perspective highlights the need for stronger leadership commitment, investment in training, and integration of accessibility into core business strategies to foster true inclusion and equality.

\bibliographystyle{ACM-Reference-Format}
\bibliography{Bib_File/references}

\section*{APPENDIX}
\label{sec:Appendix}

\subsection*{Questionnaire}

\begin{enumerate}
    \item Mention your role within the accessibility domain: \textit{(a) User Research - Accessibility}, \textit{(b) Designer - Accessibility}, \textit{(c) Accessibility Engineer}, \textit{(d) Accessibility Tester}, \textit{(e) Accessibility Subject Matter Expert / Advocate (Assisting all stakeholders to ensure accessibility)}, \textit{(f) Accessibility Customer Support}, \textit{(g) Accessibility Auditor}, \textit{(h) Accessibility Manager}, \textit{(i) Other (Specify)]}

    \item Rate your knowledge of digital accessibility (including the Web Content Accessibility Guidelines) \ding{73}\ding{73}\ding{73}\ding{73}\ding{73} \textsc{[1 being novice - 5 being expert]}

    \item How many years of experience do you have in the accessibility domain? \textit{ \textsc{[Select one:} (a) Less than 6 Months, (b) More than 6 months to 1 year, (c) 1-3 years, (d) 4-7 years, (e) 8-15 years, (f) 15+ years]}
    
    \item Your work location?
          \textit{\textsc{[Select One:}} \textit{(a) North America(US, Canada, Mexico)}, \textit{(b) South America(Brazil, Argentina, Colombia, Chile, Other SA countries)}, \textit{(c) Europe(UK, Germany, France, Italy, Spain, Other EU countries)}, \textit{(d) Asia(China, India, Japan, South Korea, Southeast Asia, Other Asian countries)}, \textit{(e) Middle East(UAE, Saudi Arabia, Israel, Other Middle Eastern countries)}, \textit{(f) Africa(South Africa, Nigeria, Kenya, Egypt, Other African countries)}, \textit{(g) Oceania(Australia, New Zealand, Other Oceanian countries)}, \textit{(h) Other (Specify)]}
          
    \item  Which of the following categories best describes your company? \textit{ \textsc{[Select one:} (a) Product-Based, (b) Service-Based, (c) Hybrid [companies that work in both product and service], (d) Other (Specify)]}

    \item Which of the following categories best describes the size of your company? \textit{\textsc{[Select One:} (a) Small scale (1-100 employees), (b) Medium scale (101-1000 employees), (c) Large scale (1000+ employees)]}

    \item What is your company's business model? \textit{\textsc{[Select One:} (a) B2B / Business-to-Business model, (b) B2C / Business-to-Con\-sumer model, (c) C2C / Consumer-to-Consumer model, (d) C2B / Consumer-to-Business model, (e) Other (Specify)]}

    \item How does your organization evaluate the accessibility of its digital products?(Select all that apply) \textit{[(a) Evaluation of the digital products with people with disabilities, (b) Automated accessibility testing tools (to test against WCAG), (c) Manual accessibility testing (to test against WCAG), (d) Third-party accessibility audits, (e) Other (Specify)]}

    \item In your organization, Is there at least one ‘accessibility champion’ in each team skilled in accessibility topics (including understanding Web Content Accessibility Guidelines)? \textit{\textsc{[Select One:} (a) Yes, (b) No, (c) Not sure]} \label{label:Q9}

    \item To what extent do you perceive leadership (engineering managers and upper management) prioritizing digital accessibility compared to other functionalities? \textit{\textsc{[Select One:} (a) Higher priority than other functional and non-functional requirements, (b) Equal Priority as that of other functional and non-functional requirements, (c)  Lower priority than other functional and non-functional requirements, (d) No Priority or consideration for accessibility, (e) Unsure]} \label{label:Q9}

    \item  How well do the organizational values and mission align with the prioritization of digital accessibility? \textit{\textsc{[Select One:} (a) Strong Alignment, (b) Alignment, (c) Neutral, (d) Misalignment, (e) Strong Misalignment]} \label{label:Q6}

    \item As a part of your job, have you conducted training sessions for designers/engineers/management on accessibility topics? \textit{\textsc{[Select One:} (a) Yes, sessions aimed to create awareness and sensitization on accessibility, (b) Yes, sessions aimed to enable the creation of accessible products by training about WCAG, Inclusive design practices, (c) No, have not conducted any training]} 

    \item What was the mode of the accessibility training sessions? \textit{\textsc{[Select One:} (a) Online - Instructor-led, (b) Online - Prerecorded videos, (c) Offline - Instructor-led, (d) Offline - Prerecorded videos played as a watch party, (e) Animated videos / Games / Simulations, (f) Other (Specify)]}

    \item What is the total duration of these trainings (per fiscal year) ? \textit{\textsc{[Select One:} (a) < 3 hours, (b) 3 to 5 hours, (c) 5 to 10 hours, (d) 10+ hours]}

    \item How effective are these training sessions in instilling a will among the participants to prioritize digital accessibility? \ding{73}\ding{73}\ding{73}\ding{73}\ding{73} \textsc{[5 = Highly effective, 1 = highly ineffective]}

    \item How effective are these training sessions in instilling skills among the participants to prioritize digital accessibility? \ding{73}\ding{73}\ding{73}\ding{73}\ding{73} \textsc{[5 = Highly effective, 1 = highly ineffective]} 

    \item In your opinion, what are the major barriers to ensuring accessibility?(Select all that apply) \textit{\textsc{[Select One or more than one:} (a) Lack of awareness about accessibility, (b) Lack of skilled developers/professionals in accessibility topics, (c) Lack of willingness to consider accessibility even though resources are offered, (d) Lack of reliable automation tools for accessibility testing, (e) Lack of adequate prioritization for accessibility (resulting in lack of time for accessibility), (f) Difficult to recruit people with disabilities with technical knowledge to evaluate product accessibility, (g) Difficult to accommodate PwDs who could provide valuable inputs on the softwares user experience, (h) Difficulty in enforcing the “shift-left” principle for accessibility, (i) Budgetary concerns to take care of accessibility, (j) Organizations feel there is no major return on investment in accessibility topics., (k) Other (Specify)]}

    \item What challenges do you encounter in adopting the “shift-left” paradigm for accessibility? [The shift-left approach emphasizes considering accessibility early in the design and development phases rather than being deferred until the later stages of development or testing stages of the product.](Select all that apply) \textit{\textsc{[Select One or more than one:} (a) Lack of Awareness, (b) Limited Engagement, (c) Resistance from Engineers, (d) Lack of Management Support, (e) Organizational Structure does not support shift-left easily, (f) Bureaucratic challenges concerning who is responsible for ensuring accessibility, (g) Other (Specify)]}

    \item Have there been initiatives or incentives implemented in your workplace to encourage engineers and management to prioritize digital accessibility features? \textit{\textsc{[Select One:} (a) Yes, (b) No, (c) In Progress]} \label{label:Q19}
          \begin{itemize}
              \item If yes, Name the initiative or describe the incentives
          \end{itemize}

    \item Which accessibility barriers do you prioritize the most in your work? \textit{\textsc{[Select One:} (a) Visual related, (b) Auditory related, (c) Cognitive related, (d) Motor related, (e) Treat all of them equally]}

   \item How frequently do accessibility SMEs collaborate with designers/engineers in addressing digital accessibility concerns? \textit{\textsc{[Select One:} (a) Very Frequently, (b) Frequently, (c) Occasionally, (d) Rarely, (e) Never]}
 
  \item To what extent do designers/engineers incorporate feedback provided by accessibility SMEs? \textit{\textsc{[Select One:} (a) Always, (b) Often, (c) Occasionally, (d) Rarely, (e) Never]}  

    \item Are there established metrics to measure the success of collaborative efforts
between accessibility professionals and engineers?
 \textit{\textsc{[Select One:} (a) Yes, (b) No, (c) In Progress]} \label{label:Q9}
  
    \item Is there a dedicated budget or resources allocated specifically for providing
digital accessibility training to engineers/designers/other stakeholders in your
organization ?
 \textit{\textsc{[Select One:} (a) Yes, (b) No, (c) Not sure]} \label{label:Q9}

    \item Is there a mandatory review for accessibility by accessibility professionals in
your organization?
 \textit{\textsc{[Select One:} (a) Yes, (b) No]} \label{label:Q9}
          \begin{itemize}
              \item If no, go to question 27
          \end{itemize}

    \item At what stage of the software development lifecycle is the accessibility review
conducted?
(Select all that apply) \textit{\textsc{[Select One or more than one:} (a) During User Research, (b) During UX Design, (c) During Software development, (d) During Software Testing]}

    \item Does your organization facilitate (by providing time and funding) accessibility
learning and encourage employees to get certifications (such as Section 508
Trusted Tester, IAAP certifications etc.)?
 \textit{\textsc{[Select One:} (a) Yes, (b) No]} \label{label:Q9}
 \end{enumerate}



\subsection*{Additional Findings}

\begin{table*}[thbp]
  \caption{Participant Profile Summary (N=160)}
  \label{tab:participantSummary}
  \centering
          \begin{tabular}{llll}
              \toprule
              \multirow{1}{*}{\textbf{Category}} & \multicolumn{1}{l}{\textbf{Sub-Category}} & \multicolumn{1}{l}{\textbf{Count}} & \multicolumn{1}{l}{\textbf{Percent}}
              \\\hline
              \multirow{3}{*}{Gender} & \multicolumn{1}{l}{Male} & \multicolumn{1}{l}{77} & \multicolumn{1}{l}{48.1\%} \\\cline{2-4}
                                   & \multicolumn{1}{l}{Female}  & \multicolumn{1}{l}{54} & \multicolumn{1}{l}{33.8\%} \\\cline{2-4}
                                   & \multicolumn{1}{l}{Not Disclosed} & \multicolumn{1}{l}{29} & \multicolumn{1}{l}{18.1\%} \\\hline
               \multirow{6}{*}{Age} & \multicolumn{1}{l}{18-24 Years} & \multicolumn{1}{l}{11} & \multicolumn{1}{l}{6.9\%} \\\cline{2-4}
                                   & \multicolumn{1}{l}{25-34 Years} & \multicolumn{1}{l}{48} & \multicolumn{1}{l}{30.0\%} \\\cline{2-4}
                                   & \multicolumn{1}{l}{35-44 Years} & \multicolumn{1}{l}{32}  & \multicolumn{1}{l}{20.0\%} \\\cline{2-4}
                                   & \multicolumn{1}{l}{45-54 Years} & \multicolumn{1}{l}{64}  & \multicolumn{1}{l}{40.0\%} \\\cline{2-4}
                                   & \multicolumn{1}{l}{55-64 Years} & \multicolumn{1}{l}{5}  & \multicolumn{1}{l}{3.1\%}\\\hline 
                 \multirow{6}{*}{Location} & \multicolumn{1}{l}{North America} & \multicolumn{1}{l}{62} & \multicolumn{1}{l}{38.8\%} \\\cline{2-4}
                                   & \multicolumn{1}{l}{Europe} & \multicolumn{1}{l}{47} & \multicolumn{1}{l}{29.3\%} \\\cline{2-4}
                                   & \multicolumn{1}{l}{Asia} & \multicolumn{1}{l}{36}  & \multicolumn{1}{l}{22.5\%} \\\cline{2-4}
                                   & \multicolumn{1}{l}{Oceania} & \multicolumn{1}{l}{8}  & \multicolumn{1}{l}{5.0\%} \\\cline{2-4}
                                   & \multicolumn{1}{l}{South America} & \multicolumn{1}{l}{5}  & \multicolumn{1}{l}{3.1\%} \\\cline{2-4}
                                   & \multicolumn{1}{l}{Middle East} & \multicolumn{1}{l}{2} & \multicolumn{1}{l}{1.2\%} \\\hline
              \multirow{4}{*}{Accessibility Experience} & \multicolumn{1}{l}{Less than 1 year} & \multicolumn{1}{l}{2} & \multicolumn{1}{l}{1.2\%} \\\cline{2-4}
                                   & \multicolumn{1}{l}{1-3 Years} & \multicolumn{1}{l}{33} & \multicolumn{1}{l}{20.6\%} \\\cline{2-4}
                                   & \multicolumn{1}{l}{4-7 Years} & \multicolumn{1}{l}{48} & \multicolumn{1}{l}{30.0\%} \\\cline{2-4}
                                   & \multicolumn{1}{l}{8-15 Years} & \multicolumn{1}{l}{42}  & \multicolumn{1}{l}{26.2\%} \\\cline{2-4}
                                   & \multicolumn{1}{l}{15+ Years} & \multicolumn{1}{l}{35} & \multicolumn{1}{l}{21.9\%} \\\hline
              \multirow{6}{*}{Domain} & \multicolumn{1}{p{.4\columnwidth}}{Accessibility Subject Matter Expert/Consultant} & \multicolumn{1}{l}{91} & \multicolumn{1}{l}{56.8\%} \\\cline{2-4}
                                   & \multicolumn{1}{l}{Accessibility Tester} & \multicolumn{1}{l}{25} & \multicolumn{1}{l}{15.6\%} \\\cline{2-4}
                                   & \multicolumn{1}{l}{Accessibility Manager} & \multicolumn{1}{l}{14} & \multicolumn{1}{l}{8.8\%} \\\cline{2-4}
                                    & \multicolumn{1}{l}{Accessibility Engineer} & \multicolumn{1}{l}{13} & \multicolumn{1}{l}{8.1\%} \\\cline{2-4}
                                     & \multicolumn{1}{l}{Designer (Accessibility Focused)} & \multicolumn{1}{l}{6} & \multicolumn{1}{l}{3.8\%} \\\cline{2-4}
                                   & \multicolumn{1}{l}{Others - Accessibility Auditor, Researcher} & \multicolumn{1}{l}{11} & \multicolumn{1}{l}{6.9\%} \\\hline                  
                                   \multirow{3}{*}{Identify as PwD} & \multicolumn{1}{l}{Yes} & \multicolumn{1}{l}{44} & \multicolumn{1}{l}{27.5\%} \\\cline{2-4}
                                   & \multicolumn{1}{l}{No} & \multicolumn{1}{l}{102} & \multicolumn{1}{l}{63.7\%} \\\cline{2-4}
                                   & \multicolumn{1}{l}{Prefer not to say} & \multicolumn{1}{l}{14} & \multicolumn{1}{l}{8.7\%}  \\\bottomrule
          \end{tabular}
  \end{table*}

\begin{table*}[ht]
\caption{Participants' Organization Summary (N=160)}\label{tab:participantOrgSummary}
\centering
        \begin{tabular}{llll}
            \toprule
            \multirow{1}{*}{\textbf{Category}} & \multicolumn{1}{l}{\textbf{Sub-Category}} & \multicolumn{1}{l}{\textbf{Count}} & \multicolumn{1}{l}{\textbf{Percent}}
            \\\hline
             \multirow{3}{*}{Company Size} & \multicolumn{1}{l}{Large scale}  & \multicolumn{1}{l}{90} & \multicolumn{1}{l}{56.2\%} \\\cline{2-4}
                                 & \multicolumn{1}{l}{Medium scale }  & \multicolumn{1}{l}{59} & \multicolumn{1}{l}{36.9\%} \\\cline{2-4}
                                 & \multicolumn{1}{l}{Small scale }  & \multicolumn{1}{l}{11} & \multicolumn{1}{l}{6.9\%}  \\\hline
            \multirow{3}{*}{Company Modal} & \multicolumn{1}{l}{B2B} & \multicolumn{1}{l}{77} & \multicolumn{1}{l}{48.1\%} \\\cline{2-4}
                                 & \multicolumn{1}{l}{B2C} & \multicolumn{1}{l}{65} & \multicolumn{1}{l}{40.6\%} \\\cline{2-4}
                                & \multicolumn{1}{l}{Others} & \multicolumn{1}{l}{18}& \multicolumn{1}{l}{11.3\%} \\\hline
            \multirow{3}{*}{Company Type} & \multicolumn{1}{l}{Service Based } & \multicolumn{1}{l}{44} & \multicolumn{1}{l}{27.5\%} \\\cline{2-4}
                                 & \multicolumn{1}{l}{Product based } & \multicolumn{1}{l}{54} & \multicolumn{1}{l}{33.7\%} \\\cline{2-4}
                                 & \multicolumn{1}{l}{Hybrid} & \multicolumn{1}{l}{52} & \multicolumn{1}{l}{32.5\%} \\\cline{2-4} 
                                & \multicolumn{1}{l}{Others} & \multicolumn{1}{l}{10} & \multicolumn{1}{l}{6.3\%} \\
              \bottomrule
        \end{tabular}
\end{table*}

\end{document}